\documentclass[apjl]{emulateapj}
\usepackage{apjfonts}

\shorttitle{Searching for AGN-driven Shocks in Galaxy Clusters}
\shortauthors{A. Cavaliere \& A. Lapi}  \journalinfo{Accepted on
ApJL.}

\begin{document}

\title{Searching for AGN-driven Shocks in Galaxy Clusters}
\author{A. Cavaliere\altaffilmark{1} and A. Lapi\altaffilmark{2,1}}
\altaffiltext{1}{Astrofisica, Dip. Fisica Univ. ``Tor Vergata'', Via
Ricerca Scientifica 1, 00133 Roma,
Italy.}\altaffiltext{2}{Astrophysics Sector, SISSA/ISAS, Via Beirut
2-4, 34014 Trieste, Italy.}

\begin{abstract}
Shocks and blastwaves are expected to be driven driven into the
intracluster medium filling galaxy groups and clusters by powerful
outbursts of active galactic nuclei or quasars in the member
galaxies; the first footprints of shock fronts have been tentatively
traced out with X-ray imaging. We show how overpressures in the
blasts behind the shock can prove the case and also provide specific
marks of the nuclear activity: its strength, its current stage, and
the nature of its prevailing output. We propose to detect these
marks with the aimed pressure probe constituted by the resolved
Sunyaev-Zel'dovich effect. We compute and discuss the outcomes to be
expected in nearby and distant sources at different stages of their
activity.
\end{abstract}


\section{Introduction}

Density jumps have been recently pinpointed by X-ray imaging of the
hot intracluster medium (ICM) that pervades galaxy groups and
clusters with average densities around $n \sim 10^{-3}$ cm$^{-3}$
and temperatures $T$ in the keV range. These jumps have been
interpreted in terms of shock fronts propagating into the ICM out to
radial distances $r \approx 0.2$ Mpc, with Mach numbers around
$\mathcal{M}\approx 1.5$, and involving energies up to $\Delta E
\approx 3\times 10^{61}$ ergs (Mazzotta et al. 2004; Mc Namara et
al. 2005; Forman et al. 2005; Nulsen et al. 2005a, b).

Shocks over large scales with such \emph{intermediate} strengths
were specifically expected by Cavaliere, Lapi \& Menci (2002) as
marks of the energy being fed back into the ICM by active galactic
nuclei (AGNs) when they flare up in member galaxies of a group or
cluster. Such events occur when a central supermassive black hole
(BH) accretes an additional mass $M_{\bullet} \sim 10^9 \,
M_{\odot}$; with standard efficiency $\eta\sim 10^{-1}$ for
mass-energy conversion, this yields over times $\Delta t\sim 10^8$
yr energies of order $2\times 10^{62}\, (M_{\bullet}/10^9\,
M_{\odot})$ ergs.

If these outputs couple at levels $f\sim$ a few percents to the
surrounding ICM, they constitute impulsive, considerable additions
$\Delta E \approx 10^{61}\, (f/5\%)\, (M_{\bullet}/10^9 M_{\odot})$
ergs to its binding energy. In fact, the latter comes to $E \approx
G\,M\, m/4\,r \approx 3 \times 10^{61}$ ergs in the central $0.2$
Mpcs of a cluster, that encompass a dark matter (DM) mass $M\sim
5\times 10^{13}\, M_{\odot}$ and an ICM mass fraction $m/M \approx
0.15$. With such appreciable ratios $\Delta E/E$, we expect in the
ICM a large-scale \textit{blastwave} bounded by a leading
\textit{shock} that starts from the host galaxy and moves into the
surrounding ICM out to several $10^{-1}$ Mpc.

In fact, the shock Mach numbers are provided in the simple form
$\mathcal{M} \approx (1 + \Delta E/E)^{1/2}$ by the hydrodynamics of
the ICM (see Lapi, Cavaliere \& Menci 2005), a good electron-proton
plasma which in thermal equilibrium (see \S~4) constitutes a single
``monoatomic" fluid with pressure $P\approx 2\, n\, kT$. For BH
masses bounded by $M_{\bullet} \la 5 \times 10^9\, M_{\odot}$
(Ferrarese 2002; Tremaine et al. 2002) relative energy inputs
$\Delta E/E\la 1$ obtain and yield just $\mathcal{M}\approx 1.5$;
they also yield standard Rankine-Hugoniot jumps of the post- to the
pre-shock density $n_2/n_1 = 4\, \mathcal{M}^2 /(\mathcal{M}^2 +
3)\approx 1.7$, consistent with the X-ray analyses. What other marks
will establish such shocks and blasts?

One is constituted by the temperature. This rises sharply across a
shock but then falls down in the blast, if nothing else by adiabatic
cooling. In any case, resolved spectroscopic measurements of the
electron $T$ require many X-ray photons, more than currently
available from distant clusters or groups.

Pressure provides another, independent mark; the electron pressure
is \emph{directly} sensed with the SZ effect (Sunyaev \& Zel'dovich
1972). The pressure at the shock is to jump up from the unperturbed
level $P_1$ by the factor $P_2/P_1 = (5\, \mathcal{M}^2 -1)/4$,
which marks a shock from a cold front; moreover, $P$ must retain
sufficiently high levels throughout the blast as to propel forward
the ICM it sweeps up.

We will see that the radial pressure run $P(r)$ actually
\emph{rises} from the leading shock to an inner ``cavity'', as long
as the blast is driven on by a central AGN. Power must be
transmitted from it to the surrounding ICM blast by means of an
intervening medium. This may be constituted either by relativistic
particles filling up a radiovolume energized by jets (Scheuer 1974;
Heinz, Reynolds, \& Begelman 1998); or by another and hotter plasma
heated up by the impact of radiation-driven superwinds (see Lamers
\& Cassinelli 1999).

Thus SZ pressure probing will have direct implications for the kind
and the time-cycle of the AGN outputs, in particular of the
radio-loud components.

\section{Computing overpressures in blastwaves}

In predicting pressure distributions, a divide is set by the source
time scale $\Delta t$ compared to the blast crossing time $R_s
/\dot{R}_s \approx 2\times 10^8\, \mathcal{M}^{-1}\, (R_s/0.2\,
\mathrm{Mpc})$ yr; here $R_s$ is the shock radial position, and
$\dot{R}_s = \mathcal{M}\, c_S$ its velocity in terms of the sound
speed $c_S= (5\,P/3\, n\, m_p)^{1/2}$.

The impact on the ICM of a short lived AGN source may be modeled as
an instantaneous central explosion launching the classic
self-similar blasts of Sedov (1959) and Parker (1963); these
propagate freely at high if decreasing Mach numbers $\mathcal{M}$
(which make $P_1$ and gravity irrelevant) into an initial density
run $n(r)$. When the latter is provided in the form $n(r)\propto
r^{-2}$ by the standard atmosphere in isothermal equilibrium, the
pressure $P$ in the blast declines toward the center. Steeper
density runs yield an even stronger decline, whilst flatter ones
still imply $P<P_2$.

But in some sources the observations recalled in \S~1 show proximity
of the shocks to the edge of the radiovolumes. This indicates blasts
being currently \emph{driven} by the relativistic particles
associated with the radiosource, and calls for considering AGN
outputs sustained during the propagation. A similar indication is
provided by the evidence at redshifts $z\ga 0.3$ of superwinds
driven by shining quasars (see Stockton et al. 2006). The moderate
values observed for $\mathcal{M}$ require including the effects of
\emph{gravity} and \emph{finite} initial pressure $P_1$.

These conditions are aptly modeled with the self-similar blasts
derived and discussed by Lapi et al. (2005). A significant testbed
is again provided by the isothermal distribution of both the DM and
the unperturbed ICM in the gravitational field of the former; their
cumulative masses out to $r$ scale like $m(<r) \propto M(<r) \propto
r$, and the ICM pressure like $P_1(r) \propto n_1(r) \propto
r^{-2}$.

The overall parameters of the perturbing blast may be derived from
the simple ``shell approximation'', long known to be effective and
precise (see Cavaliere \& Messina 1976; Ostriker \& McKee 1988).
This treats the blast as a shell containing the mass $m(<R_s)$ swept
up out to the radius $r = R_s$, and propelled outward by the
volume-averaged pressure $\langle P \rangle$ against, we add, the
upstream pressure $P_1$ and the DM gravity $G M(< R_s)\, m(<
R_s)/R^2_s$. It leads to the momentum equation
\begin{equation}
{\mathrm{d}\over \mathrm{d}t}[m(<R_s)\, v_2] = 4 \pi\, R^2_s \, [
\langle P \rangle - P_1] - {G\, M(<R_s)\over R^2_s}\, m(<R_s)
\end{equation}
in terms of the postshock speed $v_2=3\,(\mathcal{M}^2-1)\,
\dot{R}_s/4\, \mathcal{M}^2$. Here the pressure terms scale
following $P \, R^2_s =$ const, as does the gravitational term.
Self-similar solutions require also the l.h.s. to remain constant as
the blast moves out; since $m(<R_s)\,v_2\propto R_s\,\dot{R}_s-\,
R_s\,c_S^2/\dot{R}_s$ holds (with $c_S=$ const), the only consistent
solution must satisfy $\dot R_s\propto c_S$. Thus the shock moves
outward following $R_s = \mathcal{M}\, c_S \, t$.

On the other hand, the gravitational energy in the DM potential well
scales like $G \, M(<R_s)\, m(<R_s)/R_s \propto R_s$, and so do all
energies including the binding $E$; this implies for the energy
injection $\Delta E(t)\propto t$ to hold, that is, the power $L(t)
=$ const is \textit{sustained} over a crossing time. Then the blast
runs unattenuated with $\Delta E(t)/E(<R_s[t])$ and $ \mathcal{M}
\approx (1 + \Delta E/E)^{1/2}$ independent of time and position.

The actual existence of the self-similar solutions is proven from
the full hydro equations. By self-similarity these reduce to a set
of three ordinary differential equations in the normalized variable
$r/R_s$, that can be integrated inward of the shock where
Rankine-Hugoniot boundary conditions hold, see Lapi et al. (2005).
Here we focus on the shape of $P(r/R_s)$ that is plotted in detail
in Fig.~1 (\textit{left} panel).

The point to stress is that $P(r/R_s)$ \emph{rises} from the shock
to the blast inner boundary, namely, the ``piston''. This is the
contact discontinuity separating the ICM from the inner medium in
the cavity; the mass $m(<R_p)$ of swept-up gas piles up there,
causing the density to diverge mildly whereas the temperature $T(r)$
vanishes. In the blast the two combine to yield a rising $P(r/R_s)$;
at the piston this attains a value $P_p$ exceeding $P_2$ by factors
$3-4$ (see Fig.~1), larger at lower ${\mathcal M}$.

The rising trend is best understood in the shell approximation on
computing the coefficients of Eq.~(1), given the scalings above;
simple algebra and some labor yield
\begin{equation}
{\langle P \rangle\over P_2} = {4\over 5\, {\mathcal
M}^2-1}\,\left[{5\over 4}\,({\mathcal M}^2-1)+ 3\right] = {5
\,{\mathcal M}^2 + 7 \over 5 \,{\mathcal M}^2-1}~.
\end{equation}
The first additive term in the middle expression embodies the
contribution to $\langle P\rangle$ from the shell momentum; the
second, those from the initial pressure (contributing $1/3$) plus
gravitation ($2/3$). The final expression shows that $\langle
P\rangle/P_2$ and hence the monotonic $P(r)/P_2$ go to unity for
${\mathcal M} \gg 1$, the result known from Sedov blasts launched
into a density gradient with negligible gravity and initial
pressure. On the other hand, $\langle P\rangle$ and $P(r)/P_2$
increase as $\mathcal{M}\rightarrow 1$; this concurs with the
lowering jump $P_2/P_1$ to yield an overall value $P_p$ which
decreases slowly with $\mathcal{M}$, as shown by Fig.~1.

In sum, a driving source excavates in the ICM a cavity with very low
inner X-ray emission; this is bounded by the blast (from a rim at
the piston to the leading shock) in a cocoon-like topology. Enhanced
values $P_p/P_2$ relate mainly to gravity, important to the ICM as
long as all velocities do not exceed $c_S$, or the DM velocity
dispersion; so at lower $\mathcal{M}$ a relatively stronger push is
required to propel the cocoon on.

From the energy standpoint, high values of $P_p$ may be viewed as
necessary for soaking up -- in the form of work $4\, \pi \, R^2_s \,
P_p\, \dot{R}_s$ done at the piston -- the power fraction
transferred from the source via the inner medium filling the cavity.
The fraction is $\alpha \approx 1/2$ when the medium is constituted
by relativistic particles, see Cavaliere \& Messina (1976), and
$\alpha =1$ in the case of a thermalized superwind.

\begin{figure*}
\plottwo{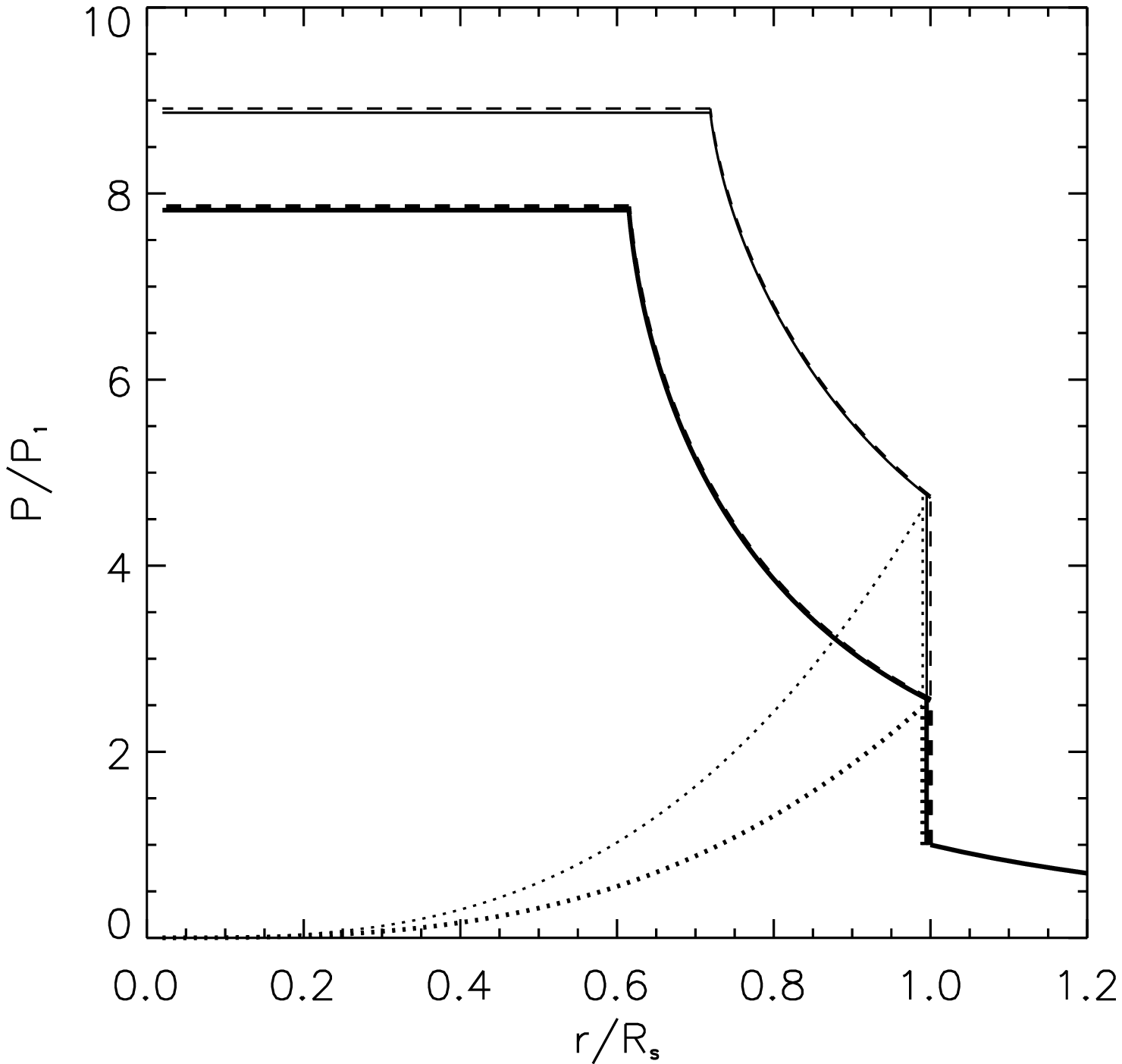}{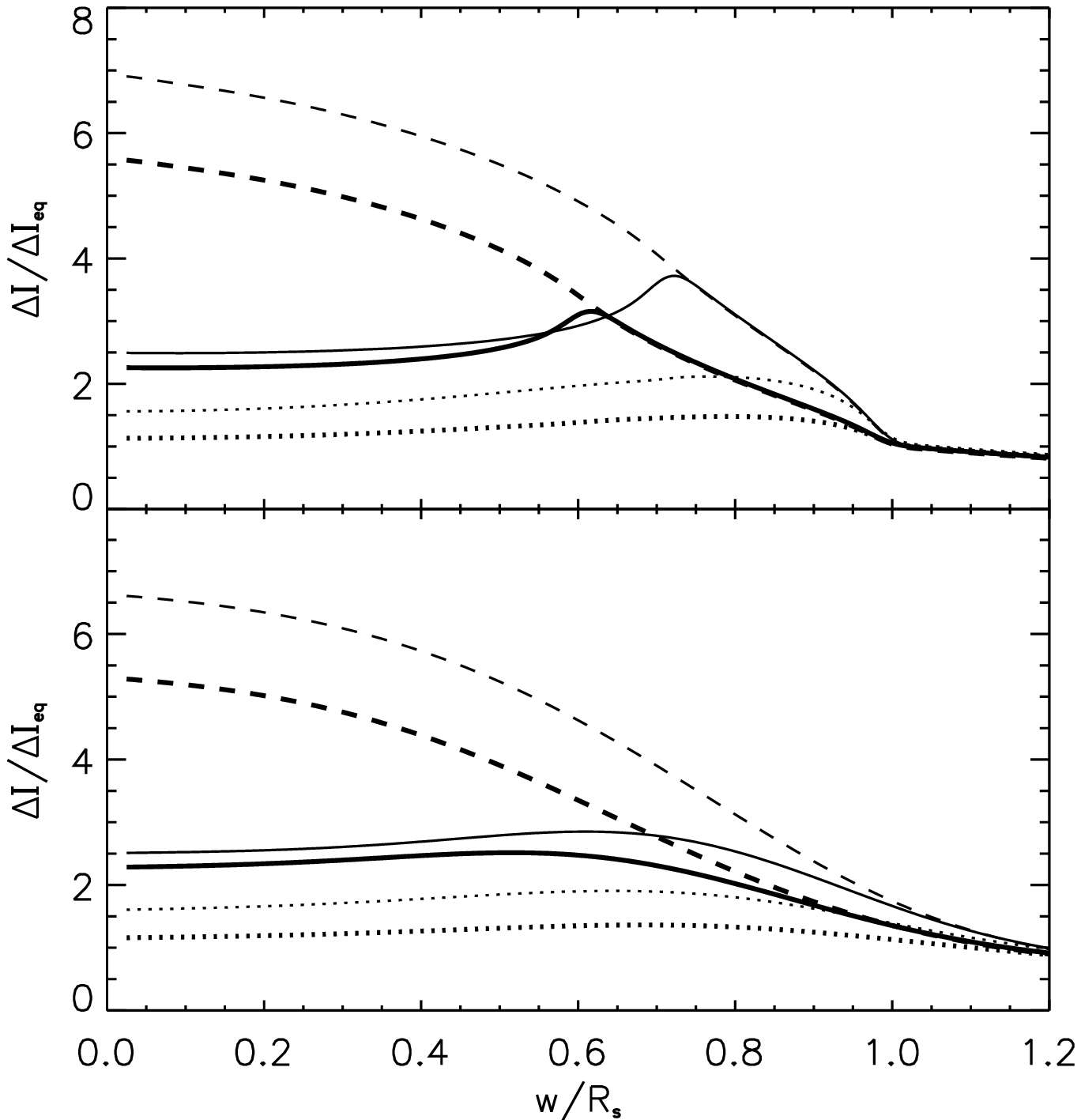} \caption{\textit{Left
panel}: overpressure in an AGN-driven blastwave; \textit{top right
panel}: enhanced SZ signal (relative to $\Delta I_{\mathrm{eq}}$ on
the l.o.s. grazing the shock) at an instrumental resolution of $1''$
in a cluster at $z=0.1$; \textit{bottom right panel}: same for a
resolution of $10''$. \textit{Thick} lines are for Mach number
$\mathcal{M}=1.5$, when the piston location is at $R_p \approx
0.62\, R_s$; and \textit{thin} lines are for $\mathcal{M}=2$, when
the piston is at $R_p \approx 0.72\, R_s$. \textit{Solid} lines are
for a blast continuously driven by relativistic particles;
\textit{dashed} lines are for a blast continuously driven by a hot
inner plasma; \textit{dotted} lines are for a standard Sedov-Taylor
blastwave launched by a sudden energy release.}
\end{figure*}

\section{Measuring overpressures with the SZ effect}

The enhanced pressures in the blast can be \emph{directly} probed
with the SZ effect. This arises when photons of the cosmic microwave
background (CMB) crossing a cluster are Compton upscattered by the
hot ICM electrons; then the black body spectrum of the CMB is tilted
slightly toward higher energies.

The resulting intensity change $\Delta I$ of the CMB radiation at a
normalized frequency $x\equiv h\nu/kT_{\mathrm{CMB}}$ is given in
the thermal, non-relativistic case (Rephaeli 1995) by
\begin{equation}
\Delta I = 2\, {(kT_{\mathrm{CMB}})^3\over (hc)^2}\,g(x)\; y ~.
\end{equation}
The effect strength is set by the Comptonization parameter
\begin{equation}
y(w) = 2\, {\sigma_T\over m_e c^2}\,
\int_0^{\ell_{\mathrm{max}}}{d\ell\, p(r)}~,
\end{equation}
just proportional to the electron \emph{pressure} $p=n\,kT$
integrated along the line of sight (l.o.s.) at a projected distance
$w$ from the cluster center; $\ell_{\mathrm{max}} = (R^2+w^2)^{1/2}$
applies if the ICM boundary is set at the virial radius $R$.

The spectral shape is encoded in the factor $g(x)$. This has a
crossover point at $\nu\approx 220$ GHz; it is positive for larger
$\nu$ with a peak at $\nu \approx 370$ GHz, and negative for lower
$\nu$ with a minimum at $\nu\approx 130$ GHz. The approximation
$g(x)\simeq -2\, x^2$ applies at the Rayleigh-Jeans end of the CMB
spectrum.

As $\Delta I\propto y\propto P$ holds, we expect that in an
AGN-driven blast the SZ signal is \emph{enhanced} from shock to
piston, relative to the equilibrium value $\Delta I_{\mathrm{eq}}$
on the l.o.s. grazing the shock. In Fig.~1 (\textit{right} panels)
we show the outcome of our computations based on Eq.~(3) and on
$P(r/R_s)$ given on the left.

We describe our results beginning with $\nu < 220$ GHz. The rise of
$\Delta I/\Delta I_{\mathrm{eq}}$ shown in the \textit{top right}
panel starts up at the shock position $w=R_s$ with a high derivative
arising from the contribution of the pressure jump $P_2/P_1$ to the
l.o.s. integral. The subsequent rise toward the piston reflects the
further pressure increase into the blast.

Inward of the piston $\Delta I/\Delta I_{\mathrm{eq}}$ rises still
\emph{further} when the inner medium is constituted by a very hot
thermal plasma with constant pressure $P_p$ (\textit{dashed} lines);
this is closely the case with a thermalized superwind, where $c_S$
is high owing to the high pressure $P_p \gg P_1$ and the low density
$n < n_1$.

On the contrary, $\Delta I/\Delta I_{\mathrm{eq}}$ \textit{declines}
for $w < R_p$ (\textit{solid} lines) when the cavity within $R_p$ is
filled with relativistic electrons (with $c_S\approx c/\sqrt{3}$).
In fact, these provide the pressure $P_p\approx 8\, P_1$ but
contribute little to the overall SZ signal on two grounds. First,
relativistic pressures $p_{\mathrm{rel}} = P_p$ imply low electron
densities even with minimal Lorentz factors of a few (see Pfrommer,
En{\ss}lin \& Sarazin 2005; Colafrancesco 2005). Second,
relativistic electrons are generally inefficient contributors at a
given $\nu$ since the tilt they cause in the CMB spectrum is
stretched toward high frequencies.

For comparison, we also show the thermal SZ effect produced by free
Sedov-Taylor blastwaves launched by a strong but short explosion
(\textit{dotted} lines). Here the pressure declines below $P_2$
after the shock jump, so $\Delta I/\Delta I_{\mathrm{eq}}$ rises
modestly inward of $w=R_s$ and soon decreases.

In fact, the \textit{right} panels of Fig.~1 show the shapes of
$\Delta I/\Delta I_{\mathrm{eq}}$ expected with SZ instruments
having two angular resolutions; in the \textit{top} panel we show
our results smoothed with a Gaussian window of $1''$ (about $2$ kpc
at $z\approx 0.1$), and in the \textit{bottom} panel with $10''$
(about $20$ kpc). The latter resolution is currently achieved, while
the former will be attained in the near future, see \S~4.

For frequencies above $220$ GHz, similar shapes obtain for $w \leq
R_s$, except that the non-thermal contribution is generally lower
than plotted in Fig.~1. The thermal/non-thermal ratio is maximized
at $\nu\approx 370$ GHz, to decline only at much larger $\nu$. At
$\nu\approx 220$ GHz only the non-thermal component survives; here
the spatial shape of the SZ signal differs significantly on two
accounts. First, the rise begins at the piston position $w=R_p$ with
a sharp transition of $p_{\mathrm{rel}}$ up to $P_p$; second,
saturation occurs at low levels.

Finally, we discuss why the initial isothermal atmosphere, where
$n\propto r^{-\omega}$ applies with $\omega=2$, provides reliable
evaluations of the SZ signals from driven blasts. Flatter runs with
$\omega < 2$ occur in a cluster at $r\la 100$ kpc; here the gravity
is weaker but the mass $m(<r)$ swept up by the blast increases
faster, and in the shell approximation Eq.~(2) yields for $<P>/P_2$
a value $6/7$ of the isothermal case. Steeper atmospheres
corresponding to $2 < \omega < 2.5$ occur for $r\ga 500$ kpc; these
also yield self-similar blasts propagating at constant $\mathcal{M}$
when acted upon by a fading AGN with integrated energy $\Delta E
\propto t^{2\,(5-2\omega)/\omega}$. The outcomes include larger
pressure jumps but also flatter runs of $P(r)$ at equal values of
$\Delta E/E$ (see Lapi et al. 2005); overall, somewhat higher levels
of $P_p$ obtain.

\section{Discussion and conclusions}

SZ signals are widely measured in clusters at levels $y\approx
10^{-4}$, see Reese et al. (2002) and Birkinshaw (2004); Lapi,
Cavaliere \& De Zotti (2003) discuss sub-arcmin resolutions to
detect the integrated effects of AGN outbursts on the ICM of
clusters and groups. Here we have focused on SZ signals
\emph{resolved} at levels of $10''$ or better, to probe the
structure and the dynamics of the shocks and blasts so produced.

The SZ probe is best used in scanning the ICM around density jumps
selected in X-rays. In fact, the bremsstrahlung surface brightness
proportional to $n^2$ is well suited for pinpointing density jumps
and providing positions and Mach numbers of candidate shocks. But
measuring $T$ from X-ray spectroscopy must contend with paucity of
photons and the narrow post-shock range where $T$ exceeds the
unperturbed value.

So in conditions of low surface brightness (outskirts or distant
structures) the SZ effect will lend a strong hand by unveiling the
other key observable, namely, the overpressures behind the shocks;
this is due to three circumstances. First, pressures are sensed
\emph{directly} through the parameter $y \propto P$. Second, $y$ is
independent of $z$ for sources wider than the instrumental
beamwidth. Third, we expect (see Fig.~1) thermal pressures to rise
throughout a blast continuously \emph{driven} over a crossing time;
from shock to piston at radii $40 - 30\%$ smaller, $P(r)$ rises up
to values $P_p\approx 8 - 9\, P_1$, considerably larger than the
shock jump $P_2/P_1\approx 2.6 - 4.8$ for $\mathcal{M} \approx 1.5 -
2$; this is due to the dynamical stress in running blasts.

This rising behavior of $P(r)$ is just opposite to the run down from
shock to center expected for \emph{free} blasts launched by
short-lived AGN activity. We compare in Fig.~1 the results we
expect; they constitute an aim particularly interesting for shocks
of intermediate $\mathcal{M}$ pinpointed in close proximity to a
radiovolume or around a currently shining AGN. Clear study cases
will be provided by clusters or groups in quiet conditions, with no
sign of outer merger induced dynamics.

What is needed for SZ probing at $z\approx 0.1$ is a resolution
around $10''$ in the upper $\mu$wave band, already approached with
the Nobeyama radiotelescope (Kitayama et al. 2004, also
\url{http://www.nro.nao.ac.jp/index-e.html/}). Upcoming instruments
such as CARMA (\url{http://www.mmarray.org/}) will do better; at
resolutions of a few arcsecs, blasts in clusters at $z\approx 0.5$
may be probed at levels comparable to the \textit{top right} panel
of Fig.~1. ALMA (\url{http://www.alma.nrao.edu/}) with its planned
resolutions around $1''$ and sensitivities down to $1\,\mu$K, will
do better yet both in the $\mu$wave and in the submm band. Following
the noise analysis by Pfrommer et al. (2005), a $5\sigma$ detection
with ALMA of the thermal SZ signal we focus on will require scanning
a limited area in the vicinity of an X-ray preselected position for
a few hours per cluster.

We add that once kpc scales will be resolved, two-fluid effects will
be interesting as may arise from disequilibrium between the
electrons and ions (see Ettori \& Fabian 1998; Fabian et al. 2006).
They may cause a gradual rise of the electron pressure in front of
the shock, to converge behind it with the declining ion pressure
toward the equilibrium values considered above.

Here we stress three issues. First, how common are shocks in
clusters? The frequent occurrence of quenched cooling flows argues
for widespread shocks driven over some Gyrs by central AGNs
injecting energies $\Delta E$ in excess of the cooling losses (see
B\^{i}rzan et al. 2004; Nulsen et al. 2005a, b).

Second, in what prevailing mode does an AGN release the driving
energy $\Delta E$, in mechanical form and radio jets or in
radiation-driven superwinds? To a first approximation the mode
little affects the total energy injected, based on the simple rule
$\Delta E \sim \alpha\, f\,\eta\, c^2 \, \dot M_{\bullet}\, \Delta t
\sim $ const that we extract from Churazov et al. (2005); that is,
mechanical energy and jets gain on grounds of coupling efficiency
($\alpha f\approx 1/2$ vs. a few $\%$) what they lose to radiation
on grounds of $\eta\dot M_{\bullet}$. Conversely, the shock
energetics alone will not distinguish the mode.

Third, what other probe may help? The specific SZ probing we propose
can trace the distinguishing features of the injection: its
timescale (past or ongoing), and its prevailing content
(relativistic particles or photons). The SZ effect resolved at
levels of $10''$ or better will directly detect the mark of a blast
launched or driven by a powerful AGN, namely, the hydro overpressure
\emph{jumping} up at the shock and \emph{sustained} throughout the
blast. But when the radiosource drive persists over the transit time
the pressure actually \emph{rises} in the blast, and the SZ signal
is enhanced; it is boosted up by plasma in the cavity if the drive
is helped by AGN superwinds.

These outcomes are independent of model details, rather they depend
on a few overall parameters: the relative injection energy $\Delta
E/E$, evaluated from $\mathcal{M}$; the active time of the AGN
compared with the blast crossing time; the injection mode, whether
dominantly in mechanical or radiative energy. We conclude that SZ
measurements concurring with the X-ray imaging can effectively probe
the injection mode, the dynamics of the blasts, and the history of
the driving AGN sources.

\begin{acknowledgements}
We thank our referee for helpful comments.
\end{acknowledgements}

\end{document}